\documentclass{article}[10pt,revtex]
\usepackage[english]{babel}
\usepackage[ansinew]{inputenc}
\usepackage[T1]{fontenc}
\usepackage{graphicx}
\usepackage{amsmath}
\usepackage{amssymb}
\usepackage{amsfonts}
\newcommand {\pspie}  {{\rm Proc. SPIE}}
\newcommand {\ao}  {{\rm Appl. Op.}}

\begin{document}

\title{X-ray propagation through hollow channel: \\
PolyCAD - a ray tracing code (1)}
\date{}

\author{D. Hampai$\,^{1}$, S.B. Dabagov$\,^{2,3}$, G. Cappuccio$\,^{1,2}$ and G. Cibin$\,^{2,4}$}
\maketitle
 {\centering {$^{1)}$ CNR - ISMN, Via Salaria Km 29,300,
I-00016 Monterotondo Scalo, Rome, Italy}\\
\centering {$^{2)}$ INFN - LNF, via E. Fermi, 40, I-00044, Frascati, Rome, Italy}\\
\centering {$^{3)}$ RAS - P.N. Lebedev Physics Institute, 119991
Moscow, Russia} \\
\centering {$^{4)}$ IMONT, P.za dei Caprettari, 70, I-00186, Rome,
Italy}}

\begin{abstract}
A new CAD program, PolyCAD, designed for X-ray photon tracing in
polycapillary optics is described. To understand the PolyCAD code
and its results, the theoretical basis of X-ray transmission by a
single cylindrical channel (monocapillary) is discussed first.
Then the simplest cases of cylindrically and conically shaped
polycapillary optics are examined. PolyCAD allows any type of
X-ray source to be used: an X-ray tube of finite beam dimensions
or an astrophysical object can be simulated in combination with
the polycapillary optics. The radiation distribution images formed
on a screen located at various focal distances are discussed. The
good agreement of some of the PolyCAD results with those reported
in earlier papers validate the code. This is the first paper of a
series dedicated to the development of an exhaustive CAD program,
work is in progress to develop the code to include other
polycapillary-optics shapes, such as semi-lenses and full-lenses.
\end{abstract}

{\bf OCIS: 340.0340, 340.7470, 080.2740, 230.7380, 999.9999.}

\section{Introduction}
\rm
 The use of polycapillary optics to control X-ray beams in
analytical X-ray apparatus for diffraction and fluorescence
analyses is becoming increasingly important. In the near future
polycapillary optics will be widely used in many different fields,
e.g., aero-space research, medicine, biology, and so
on\,\cite{vvaa}\,. Our strong interest in the development of these
devices led to the idea of creating a CAD program that allows one
to simulate the process of radiation propagation through
polycapillary optical systems and to visualize the radiation
distributions at the optical output. At present there are very few
procedures available for evaluating X-ray transmission by
capillary structures in the case of peculiar configurations, see
for instance\,\cite{Furuta1, Chen1, Hoffman1, Thiel1, Xiao1,
Guan1, Vincze1}\,. One of the first papers\,\cite{Furuta1}
reported coherent radiation transmission by a hollow glass pipe.
This code also considered i) the presence of a rough surface as an
X-ray anomalous dispersion effect, ii) radiation penetration into
the channel wall, and iii) the possible presence of micro-dust
inside the channel.

Another algorithm\,\cite{Chen1} for X-ray transmission by
capillaries of various shapes runs in the geometrical optics
approximation; for this reason is simpler and more flexible than
the previous model.

The first X-ray tracing codes were developed in
1992\,\cite{Hoffman1,Thiel1} and by the Institute for Roentgen
Optics, for a review see\,\cite{optics}\,. However, these
simulations used a number of simplifications based either on the
cylindrical symmetry for a shaped capillary system or on the
meridional ray approximation, which is valid only for describing
radiation propagation through a blend channel or a monocapillary
concentrator.

To the authors' knowledge, the most advanced and complete
softwares for X-ray tracing inside capillary channels to date are
three: the first code\,\cite{Xiao1,Xiao2} traces the trajectory of
each photon including the corrections of absorption and roughness;
in according with some experimental results, the authors could
obtain the roughness of the channel. The second one\,\cite{Guan1}
uses the SHADOW ray tracing software, adapting it to channel
shape.

The last code\,\cite{Vincze1,Vincze2} uses a Monte Carlo
simulation for X-ray radiation propagation through hollow
channels. The theoretical results obtained by this code agree
quite well with the experimental data, although the algorithm is
rather simple due to its geometrical optics approximation and to
the circular cross section of the channel shape.

The main aim of practically all the simulation codes is to optimize both the channel size and the optical shape in order to obtain highly efficient optical systems. Obviously, this is very important from the viewpoint of the development of capillary optics technology. However, analysis of the radiation distribution features, reported in a number of papers\,\cite{Sultan1, Sultan2, Sultan3} is particularly interesting. (Some of the earlier publications are cited in reference \,\cite{Sultan3}). In these papers the experimental data were validated by means of analytical estimations based on the wave theory of radiation propagation.

Before we continue any further, let us clarify the terminology used in this work: 1) a monocapillary is a single capillary; 2) a polycapillary is a set of closely and packed monocapillaries; 3) a lens is a device that concentrate the radiation in a point or a small region; 4) a semi-lens, or a half-lens is a device that can concentrate a quasi-parallel beam in a point and vice versa in the reverse geometry.

Here, we introduce a new X-ray tracing code for polycapillary optics, named \emph{PolyCAD}. A previous version was designed for cylindrical optics only\,\cite{Hampai1, Hampai2}\,. Now, the software can simulate monocapillary and polycapillary optics with cylindrical and conical shapes, while work is in progress to include lenses and semi-lenses. The advantage of the code lies in its precise mathematical solutions for each given optical shape. Comparison of the results of PolyCAD and of the previous algorithms\,\cite{Thiel1, Xiao1, Guan1, Vincze1, optics} revealed some differences, due to the fact that PolyCAD is free of many of the algorithmic constraints.

In the first part of our paper, the theoretical basis of a cylindrical monocapillary is explained; in the second, the numerical results and the simulations are reported both for cylindrical and for conical mono- and polycapillaries, using a point-source and an extended-sources, like a conventional X-ray tube.

\section{Theoretical ground}

Thanks to some simple formulas it is possible to estimate the correct behavior of the rays inside the capillary. The total intensity $I(\theta)$, evaluated at the angle $\theta$ behind a capillary, is given by all the contributions of the photon families that pass through the capillary; each family is defined by the number of its reflections:
\begin{equation}
I(\theta)\,=\,I_{0}(\theta)+I_{1}(\theta)+\cdots+I_{N}(\theta)
\end{equation}
where $N$ is the maximum number of reflections.

It is clear that the intensity $I_{m}(\theta)$ is strictly dependent on the reflection coefficient $R_{0}$, namely, on the $m^{th}$ power of $R_{0}$:
\begin{subequations}
\begin{align}
I_{0}(\theta)\,=\,R_{0} & \,\theta\in(-\theta_{0}\,,\,\theta_{0})\\
\vdots\qquad & \,\nonumber\\
I_{N}(\theta)\,\propto\,R_{0}^{N} & \,\theta\in(-\theta_{N}\,,\,\theta_{N})
\end{align}
\end{subequations}

This contribution is drastically small if the number of reflections is high, because $R_{0}$ is always less than one, (Example - $R_{0}=\left\{ 0.99, \,0.8, \,0.7 \right\} $ $\longrightarrow R_{0}^{50}=\left\{ 0.61, \,1.42\times10^{-5}, \,1.80\times10^{-8} \right\} $).

In order to consider the total radiation intensity on a screen behind a monocapillary, $I(\theta)$ has to be integrated over the whole plane, so:
\begin{equation}
I_{tot}(\theta)\,=\,\int\int d\phi\,d\theta\,\frac{dI}{d\theta}\, \propto \, \int d\phi \left\{ \sum^{N}_{i=0} \, I_{i}(\theta) \right\} \, = \, 2\,\pi \, \left\{ \sum^{N}_{i=0}\,I_{i}(\theta) \right\}
\label{Intensity}
\end{equation}
where the last mathematical passage is possible due to the axial symmetry of the system.

As can be seen from Fig. \ref{fig_ref_ter}, if the screen is placed at $P1$, near the focal point $f2$, characterized by the ray family with a maximum number of two reflections, the image on the screen is smaller than in any other configuration. Moreover, for only two reflections, there are five different areas of interaction (Figs. \ref{fig_ref_ter} and \ref{mono_f}). However, if the screen is placed at $P2$, far from the focal point, the picture on the screen does not have a simple shape, but has some zones that are more intense than others (Figs. \ref{fig_ref_ter} and \ref{mono_3}).

Now that the radiation intensity has been defined (Eq. \ref{Intensity}), the inner surface of the capillary can be divided into a discrete set of zones according to the number (1, 2,\dots) of reflections of each photon trajectory. If the origin of the reference system $(x, y, z)$ is located at the end of the capillary and the z-axis coincides with the capillary axis, each zone along the surface has a length given by the following equation:
\begin{equation}
z(\theta)\,\simeq\,f\times I(\theta),\qquad\left( \theta\ll1\right)
\end{equation}
where $f$ is the distance between the end of the capillary and the screen. Looking at Fig. \ref{fig_ref} and considering only the upper propagation case, each area has a maximum angle $\theta$ and a portion of interaction on the screen, defined by:
\begin{subequations}
\begin{eqnarray}
z_0(\theta) & & \rightarrow \qquad (0,\,\frac{d}{2}+f\theta_0) \qquad \qquad \qquad \qquad 1\\
z_1(\theta) & & \rightarrow \qquad (-\frac{d}{2}+f\theta_0,\,\frac{d}{2}+f\theta_1) \qquad \qquad   R_0\\
z_2(\theta) & & \rightarrow \qquad (-\frac{d}{2}+f\theta_1,\,\frac{d}{2}+f\theta_2) \qquad \qquad   R_0^2\\
\vdots & & \nonumber\\
z_m(\theta) & & \rightarrow \qquad (-\frac{d}{2}+f\theta_{m-1},\,\frac{d}{2}+f\theta_m) \qquad          \, \, \, R_0^m
\end{eqnarray}
\end{subequations}
where $d$ is the diameter of the capillary and $R_0^m$ is the $m^{th}$ reflection coefficient. The maximum angle for each reflection family can be evaluated as:
\begin{subequations}
\begin{eqnarray}
\theta_0 \,= \frac{1}{2}\,\frac{d}{s+L} \qquad && \\
\theta_1 \,= \frac{3}{2}\,\frac{d}{s+L} \qquad && \Delta\theta_{1,\,0}\,=\frac{d}{s+L} \\
\theta_2 \,= \frac{5}{2}\,\frac{d}{s+L} \qquad && \Delta\theta_{2,\,1}\,=\frac{d}{s+L} \\
\vdots \qquad \qquad \qquad & & \nonumber\\
\theta_m \,= \frac{2m+1}{2}\,\frac{d}{s+L} && \Delta\theta_{m,\,m-1}\,=\frac{d}{s+L}
\end{eqnarray}
\end{subequations}

From these formulas it is clear that the difference between two consecutive maximum angles is a constant and depends only on the properties of the system. When these properties are fixed, the maximum number of reflections is simple to evaluate:
\begin{eqnarray}
N&=&\frac{\theta_c-\theta}{\Delta\theta}\,=\,\frac{(d/2s)-d/[2(s+L)]}{d/(s+L)}\,= \nonumber \\
\,&=&\frac{1}{2}\left(\frac{s+L}{s}\,-1 \right)\,=\,\frac{L}{2s}
\label{N_max_1}
\end{eqnarray}

Since there must be at least one reflection, Eq. \ref{N_max_1} becomes the well known expression:
\begin{equation}
N\,=\,\left[\,\frac{L}{2s}\,\right]\,+\,1
\label{numb}
\end{equation}

Equation \ref{numb} states that the maximum number of reflections depends only on the length and on the diameter of the capillary; this result is exact if the source stays along the monocapillary z-axis. Equation \ref{numb} also remains a good approximation if the source is close to the optical axis (Fig. \ref{mono_off}). Obviously, in the general case, there are some problems because an off-axis source makes some zones inside the capillary unsuitable for photon reflection\,\cite{Kumakhov1}\,.

\section{The code}

In our previous paper\,\cite{Hampai1} the code was based only on the geometrical symmetry of the system, so the radiation image on a screen could be evaluated only if the shape of capillary was cylindrical. This new version of PolyCAD is completely different from the old one.

While upgrading the code to include other capillary shapes, the complicated definition of the reflection angle $\alpha_{ref}$ introduced some problems. For a cylindrical capillary it is easy to find the following relation for $\alpha_{ref}$:
\begin{equation}
\alpha_{ref}\,=\,\arccos{\left(\frac{\cos{\alpha}}{\cos{\omega}}\right)}\,=\,\arccos{\left\{\frac{\cos{\alpha}}{\cos{\left[
\arctan{\,\left( \tan{\alpha}\times\sin{\theta}\right)}
\right]}}\right\}}
\label{alpha_ref_cyl}
\end{equation}
where $\alpha$ is the zenithal angle and $\theta$ is the angle on the inlet plane (see Fig. \ref{geometry} and Ref.\,\cite{Hampai1}).

In the \emph{conical case}, the $\omega$ angle, formed by the two projections on the tangent plane of the X-ray photon trajectory and of the z-axis, is still a function of $\alpha$ and $\theta$, but now there is also an indirect dependence on the conical semi-opening angle $\beta$. This indirect dependence is due to the fact that the trajectory of a photon is generally oblique with respect to the capillary axis. Thus, a new parameter that takes into account the off-axis angle must be included, which does not allow us to transfer in a simple way a 3-D problem into a 2-D one, as done in the case of cylindrical optics\,\cite{Hampai1}\,. Nevertheless, by knowing the photon starting point, i.e., the source point $P_s$ and the inlet point $P_0$, it is possible to define the vector-direction of the photon. An equation system between the photon path and the capillary surface equation makes it possible to describe the trajectory of the photons inside any generic capillary optics.

\section{Numerical results}

To show how PolyCAD works, we report the most significant results. We would like to emphasize that PolyCAD can simulate any kind of source, from a point source, located at finite or infinite distance, to an extended source, and the latter can also have a 3-D shape.

\subsection{Cylindrical monocapillary}

We first considered a monocapillary or polycapillary with cylindrical channels. Even though the system geometry in this case is so simple that the results can be calculated manually, to check that the software worked well the results were compared with those of the old version of PolyCAD designed just for cylindrical capillaries\,\cite{Hampai1, Hampai2}\,.

As a preliminary step we considered a point source located at finite distance. The amplitude of the incident angle $\alpha$ can be chosen randomly from zero to any prefixed value. Obviously, in the best case this angle is equal to the critical angle, $\alpha=\theta_{c}$. Moreover, each X-ray photon will have an inlet angle $\delta$ in the $I_{p}$ plane that is strictly connected with the intersection point $P_{0}\,(x_{0},\,y_{0})$ (Fig. \ref{geometry}).

To simplify the analysis of the radiation distribution behind the capillary system, the image shape was considered in two different screen positions: i) one placed at the focus f2 of the monocapillary or polycapillary and ii) the other beyond the f2 position (Fig. \ref{fig_ref_ter}).

To compare these calculations with the results obtained by Dabagov and Marcelli\,\cite{Sultan1}\,, the length of the \emph{cylindrical monocapillary} was chosen such as to provide at least a double reflection mode of propagation. The final results obtained are illustrated in Figs. \ref{mono_f}, \ref{mono_3} and \ref{poli_f}.

The conditions for the simulation are: i) a 1 keV point source and ii) a cylindrical monocapillary 10 cm long with a radius of $\rho=10^{-1}$ cm. In Figs. \ref{mono_f} and \ref{mono_f_ist} a single central spot is present and four rings formed by the convergent rays related to the different reflection numbers (compare Figs. \ref{mono_f} and \ref{fig_ref_ter}).

In Figs. \ref{mono_3} and \ref{mono_3_ist} many sharp rings are present. The central ring is given by the X-ray photons that cross the monocapillary without interacting with the surface. The second is formed by X-ray photons that hit, once at worst, the monocapillary inner surface. The third ring is due to X-rays that produce only one reflection with the channel. The fourth ring is due to X-rays that make one or two reflections. Finally the outer ring is formed by ray vectors that undergo only two reflections.

\subsection{Cylindrical polycapillary}

In the case of a \emph{cylindrical polycapillary} the simulation parameters are i) a 1 keV point source; ii) a polycapillary 10 cm long with a radius of $\rho=1$ cm; iii) a single-channel radius of $10^{-3}$ cm.

Looking at Fig. \ref{poli_f}, for this configuration it is possible to observe a slightly bigger spot in the center, while the point density distribution of the halo decreases going from the center to the periphery. This last effect is due to the fact that each X-ray photon interacts many times with the channel surface and the number of interactions increases going from the center to the periphery.

\subsection{Conical monocapillary}

As the next step we will consider a conical capillary with an X-ray point source located at finite distance. Even though the amplitude of the incident angle $\alpha$ can be chosen in a random way from zero to any prefixed value, in order to maximize the number of incoming photons that undergo total external reflection the following relation should be satisfied: $\alpha + \beta = \theta_{c}$, where $\beta$ is the semi-opening angle of the cone.

It is not a simple job to define the focal plane for a conical capillary because each photon after each interaction with the channel surface undergoes a number of reflections with increasing incidence angular values: $\alpha + 2\,\beta$, $\alpha + 4\,\beta$, \ldots etc., according to the different directions of the incident rays. Taking this into account we decided to choose the position of the focal plane by minimizing the divergence of the exit beam. To compare the results of the conical and cylindrical capillaries, we again chose X-ray photons that undergo at least two reflections.

The parameters of the conical monocapillary are similar to the cylindrical, i.e., i) the capillary length is 10 cm; ii) the entrance radius $\rho_i=10^{-1}$ cm, while the exit radius is $\rho_f=0.8 \times 10^{-1}$ cm; iii) the energy of the photon is still 1 keV.

As for the cylindrical capillary, simulations of the focal and out-of-focus radiation distributions are shown in Figs. \ref{mono_con_f}, \ref{mono_con_3} for the \emph{conical monocapillary}.

\subsection{Conical polycapillary}

For a \emph{conical polycapillary} the simulation results are shown in Fig. \ref{poli_con}. The parameters are: i) a 1 keV point source; ii) a polycapillary 10 cm long with the entering radius of $\rho_i=1$ cm, while the exit radius is $\rho_f=0.8$ cm; iii) a single-channel radius of $10^{-3}$ cm and $0.8 \times 10^{-3}$ cm, respectively for the entering and the exit plane.

In a following paper we will discuss in more detail the differences between conical and cylindrical capillaries. In particular we will show that the conical capillary has a more intense central spot and also a bigger halo.

\subsection{Cylindrical monocapillary, off axis case}

In this section, we discuss the problem of an X-ray source located off-axis. The point source can be located at finite or at infinite distance from the capillary, so the behavior of a \emph{cylindrical monocapillary} system differs accordingly (Figs. \ref{mono_off} and \ref{mono_para}).

Figure \ref{mono_off} shows a simulation of a cylindrical monocapillary with a point source that is not placed along the z-axis, with coordinates: $P_s=(0.05,0.05,3.3)$ cm. The other physical parameters are capillary length = 10 cm, capillary radius = $10^{-1}$ cm.

We would like to point out that in this configuration the image has a symmetric shape with respect to an ideal line, which has a $45^{\circ}$ angular inclination due to the mirror-like behavior of each half of the channel. A large circular halo is also present, due to all the photons that cross the capillary without any interaction.

In Fig. \ref{mono_para} the situation is completely different because the point source is again located off-axis but at an infinite distance from the capillary. The zenithal and equatorial angles of the beam are respectively $(\alpha=1.3^{\circ},\, \delta=45^{\circ})$, while the channel parameters are, as usual, length = 10 cm, radius = $10^{-1}$ cm. This configuration means that some parts of the internal channel surface cannot be reached by X-ray photons, so now certain regions in the image are completely empty.

It is essential to emphasize that the zenithal angle is so large that many regions of the capillary wall will not be active for X-ray propagation.

\subsection{Extended sources}

The power of PolyCAD program is that it can deal with any geometrical and optical configuration. It means that we may easily simulate the behavior of a source of finite dimensions like, for example, conventional X-ray tube. In order to treat this situation an offset for a single point-source is introduced in the program. In such a mode a finite area source can be simulated by random distribution of points inside a specific area. As practical application let to consider a `Long Fine Focus' of X-ray tube (for instance, a Cu anode, $8\,keV$), where the electron spot on the anticathode has the dimension of $0.4\times12\,mm$. Such kind of tube allows us to have both an `optical point-focus' (Fig. \ref{point}) and an `optical line-focus' (Fig. \ref{line}), where in these simulations we used \emph{cylindrical polycapillary optics}. The source dimensions are respectively $0.4\times1.2\,mm$ and $0.04\times12\,mm$.

\section{Conclusion}

In this paper we have reported the first part of a \emph{PolyCAD} program designed for capillary optics. Using the ray optics approximation in the ideal case of total external reflection, i.e., without absorption effects by the optical channel walls, this algorithm allows us to i) simulate the passage of the X-ray beam inside the capillary channels for sources of different shapes, ii) visualize the spot images formed on a screen at different distances from the optics.

After a description of the theoretical basis and of the computational details, we reported the numerical analysis for various source-optics configurations. In the case of a cylindrical monocapillary, we found good agreement with previous results\,\cite{Sultan1}\,, both for single and for multiple reflection modes.

We would like to point out that this code can accurately describe the radiation distribution behind the optics, as shown in Figs. \ref{mono_f_ist} and \ref{mono_3_ist}. When the source is located along the axis of the capillary, and in particular taking into account a single channel, photons are present in all the image areas; however it is clear that by changing the source-optics conditions, the photons could be absent in some areas.

At present we are working on increasing the \emph{PolyCAD} program capabilities to deal with other polycapillary-optics shapes, such as semi-lenses and full-lenses. However, to understand the fine features of X-ray propagation through polycapillary lenses it is mandatory to consider the X-ray wave interaction with the inner capillary surface (see details in the review\,\cite{Sultan3}). This will constitute the future development of \emph{PolyCAD}.

\section{Acknowledgment}

We are grateful to M.A. Kumakhov for his continuous interest and support. This work was done within the frame of the FSSRIS Project `Multipurpose Innovative Plants for the UV and X-ray Production' (CNR - MIUR, Legge 449/97), and was partly supported by the POLYX project (Group V, LNF - INFN).


\newpage

\section*{List of Figure Captions}

\begin{figure}[htbp]
\begin{center}
\includegraphics[width=8.4 cm]{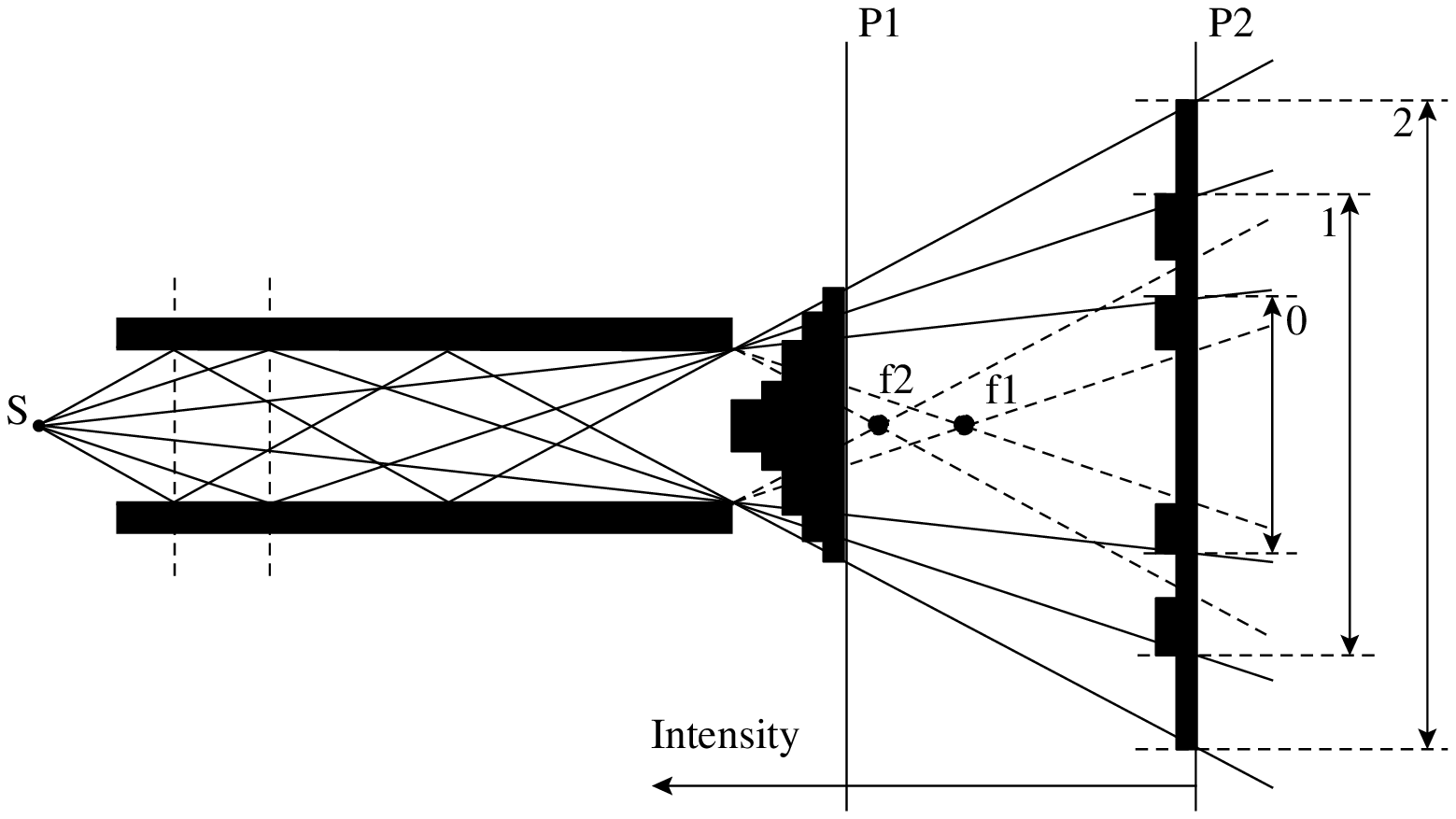}
\caption{}
\label{fig_ref_ter}%
\end{center}
\end{figure}

\noindent Fig. \ref{fig_ref_ter}. Cross section of cylindrical monocapillary with on-axis source. The expected beam intensity on a screen located in two different positions is shown. The image shape changes according to position, but note that areas without any counts are not allowed. Focal points $f1$ and $f2$ are due to rays that undergo one and two reflections, respectively.

\begin{figure}[htbp]
\begin{center}
\includegraphics[width=8.4 cm]{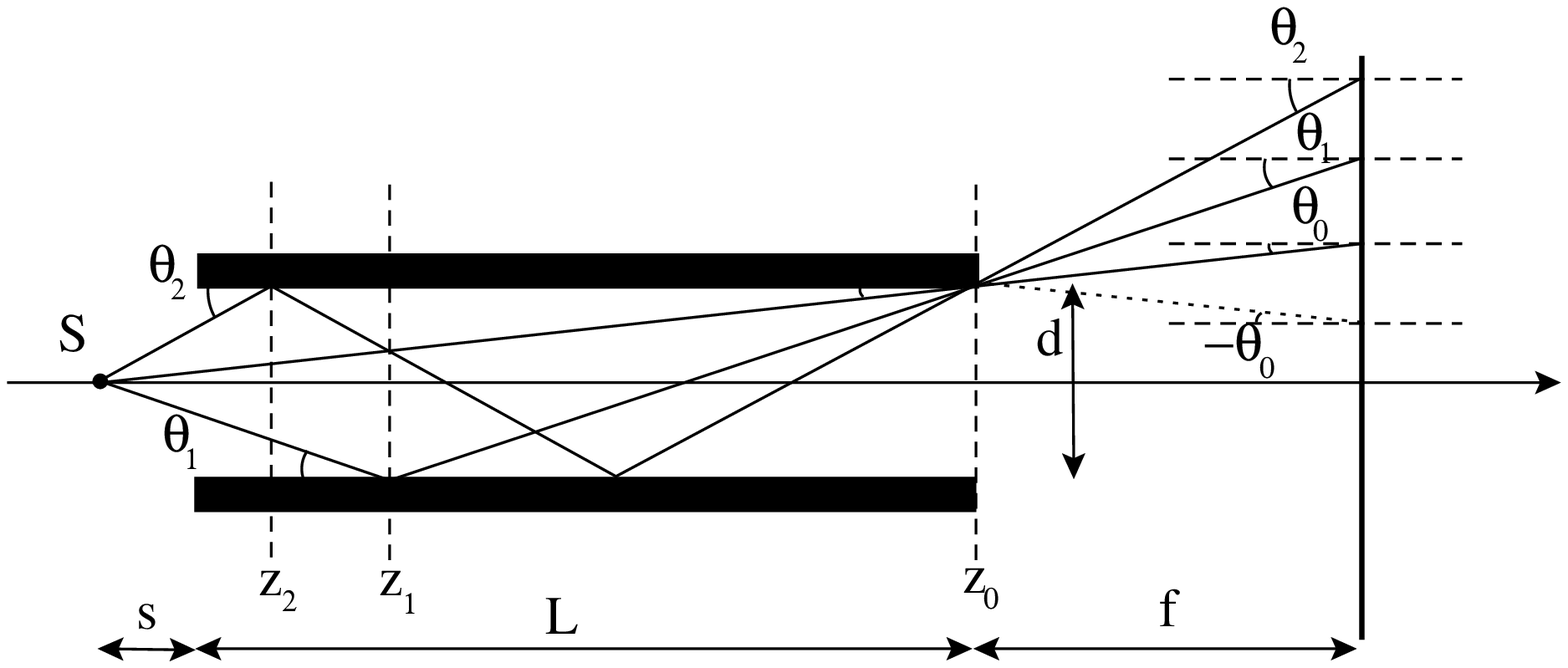}
\caption{}
\label{fig_ref}%
\end{center}
\end{figure}

\noindent Fig. \ref{fig_ref}. Length of capillary $L$; distance from source $s$; distance from screen $f$; channel radius $d$. The maximum angles allowed for each ray family are defined. Rays entering between $z_0$ and $z_1$ undergo only one reflection, rays between $z_1$ and $z_2$ undergo only two, and so on.

\begin{figure}[htbp]
\begin{center}
\includegraphics[width=8.4 cm]{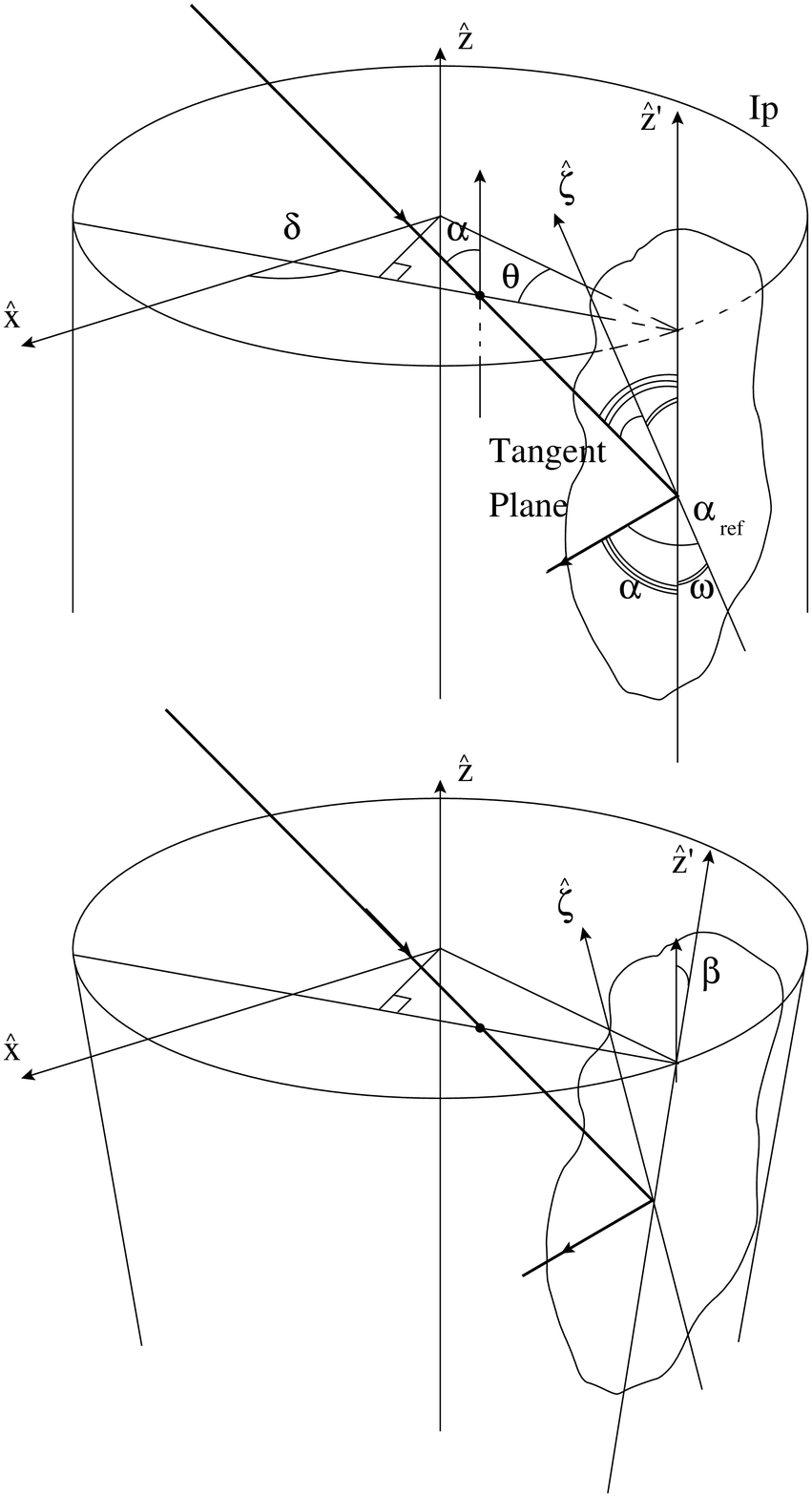}
\caption{}
\label{geometry}%
\end{center}
\end{figure}

\noindent Fig. \ref{geometry}. Three-dimensional drawing of photon reflection from a cylindrical and a conical capillary wall. Bold line shows photon path. $I_{p}$ is the transverse cross section plane at the capillary entrance. See text for symbol meaning.

\begin{figure}[htbp]
\begin{center}
\includegraphics[width=8.4 cm]{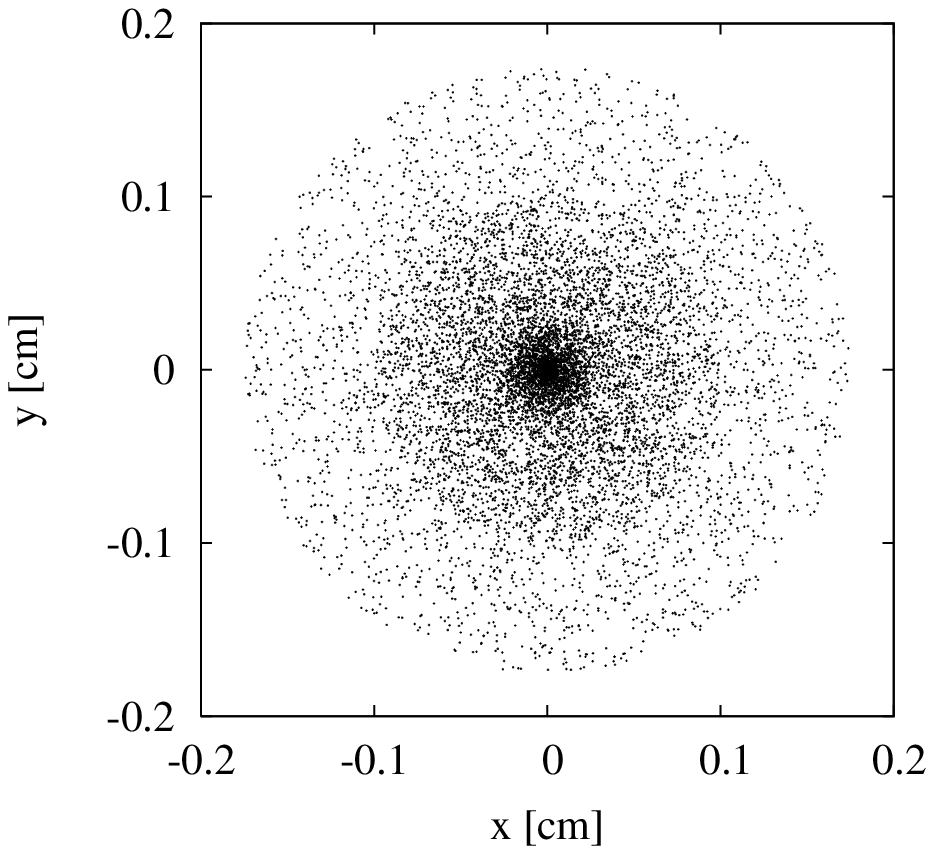}
\caption{}
\label{mono_f}%
\end{center}
\end{figure}

\noindent Fig. \ref{mono_f}. \emph{Cylindrical monocapillary}. Image of a point source on a screen located at $f2$ (Cf. Fig. \ref{fig_ref_ter}. The intense central spot is formed by all the ray families. Channel dimensions are radius = 0.1 cm and length L = 10 cm. As per theory there are no empty areas in the image.

\begin{figure}[htbp]
\begin{center}
\includegraphics[width=8.4 cm]{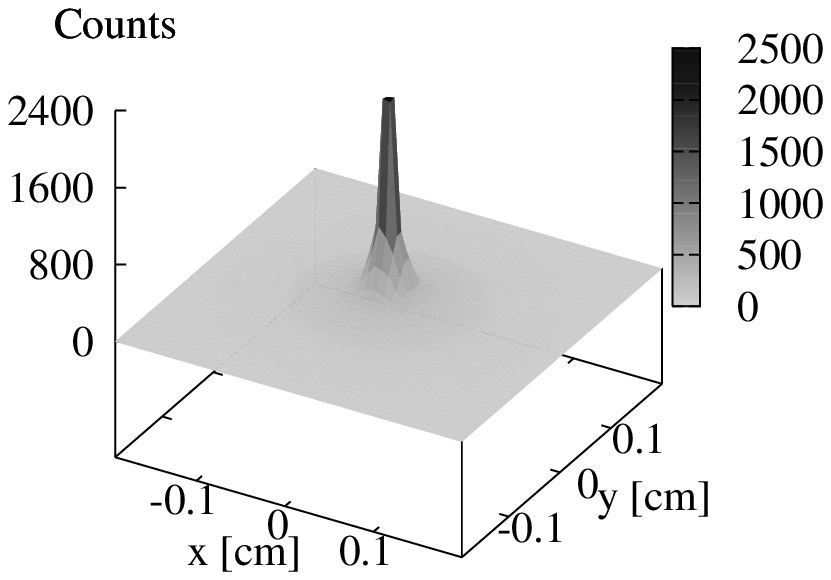}
\caption{}
\label{mono_f_ist}
\end{center}
\end{figure}

\noindent Fig. \ref{mono_f_ist}. Three-dimensional density distribution of the photons in Fig. \ref{mono_f}. Source and screen are in focal points $s$ and $f2$, respectively. The radiation distribution inside the spot is never equal to zero.

\begin{figure}[htbp]
\begin{center}
\includegraphics[width=8.4 cm]{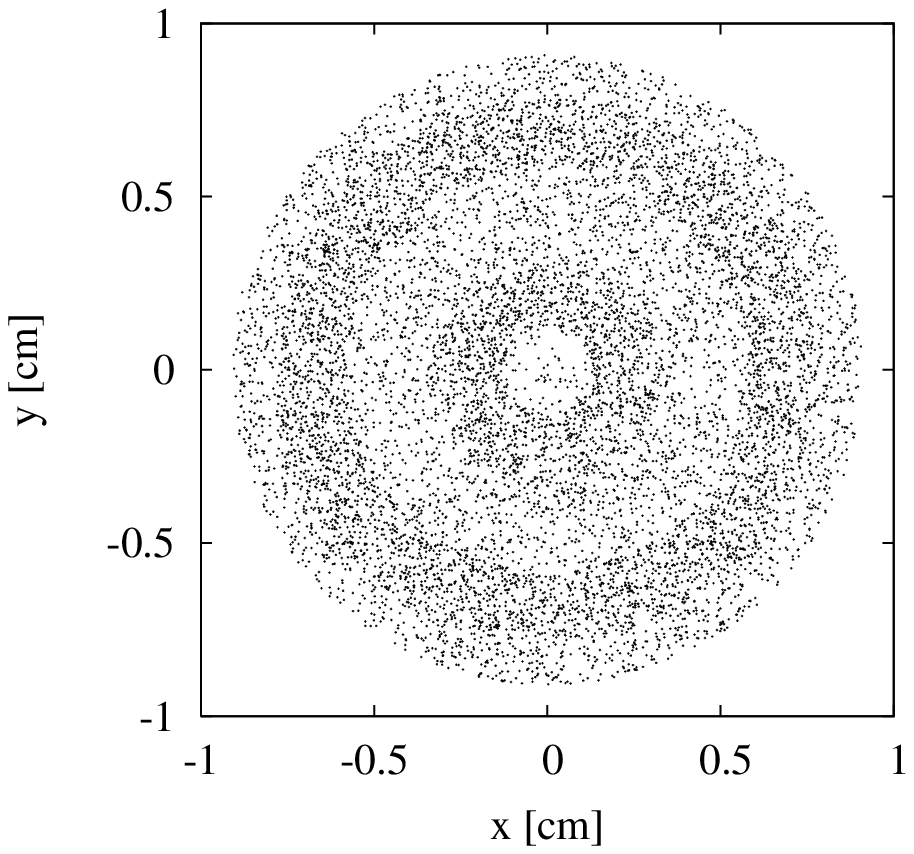}
\caption{}
\label{mono_3}%
\end{center}
\end{figure}

\noindent Fig. \ref{mono_3}. Point-source image on a screen located at $P2$ (Cf. Fig. \ref{fig_ref_ter}) out of focus behind the \emph{cylindrical channel}. The distribution consists of concentric rings of different intensities with no empty areas.

\begin{figure}[htbp]
\begin{center}
\includegraphics[width=8.4 cm]{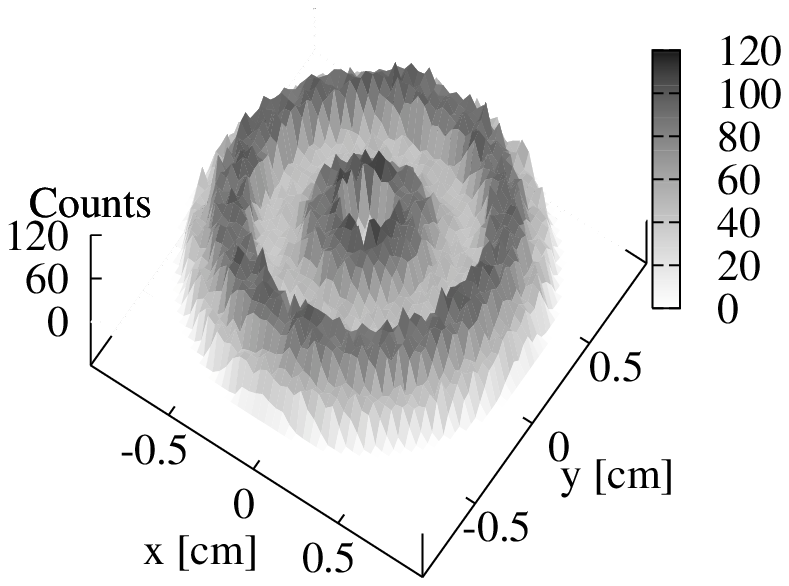}
\includegraphics[width=10 cm]{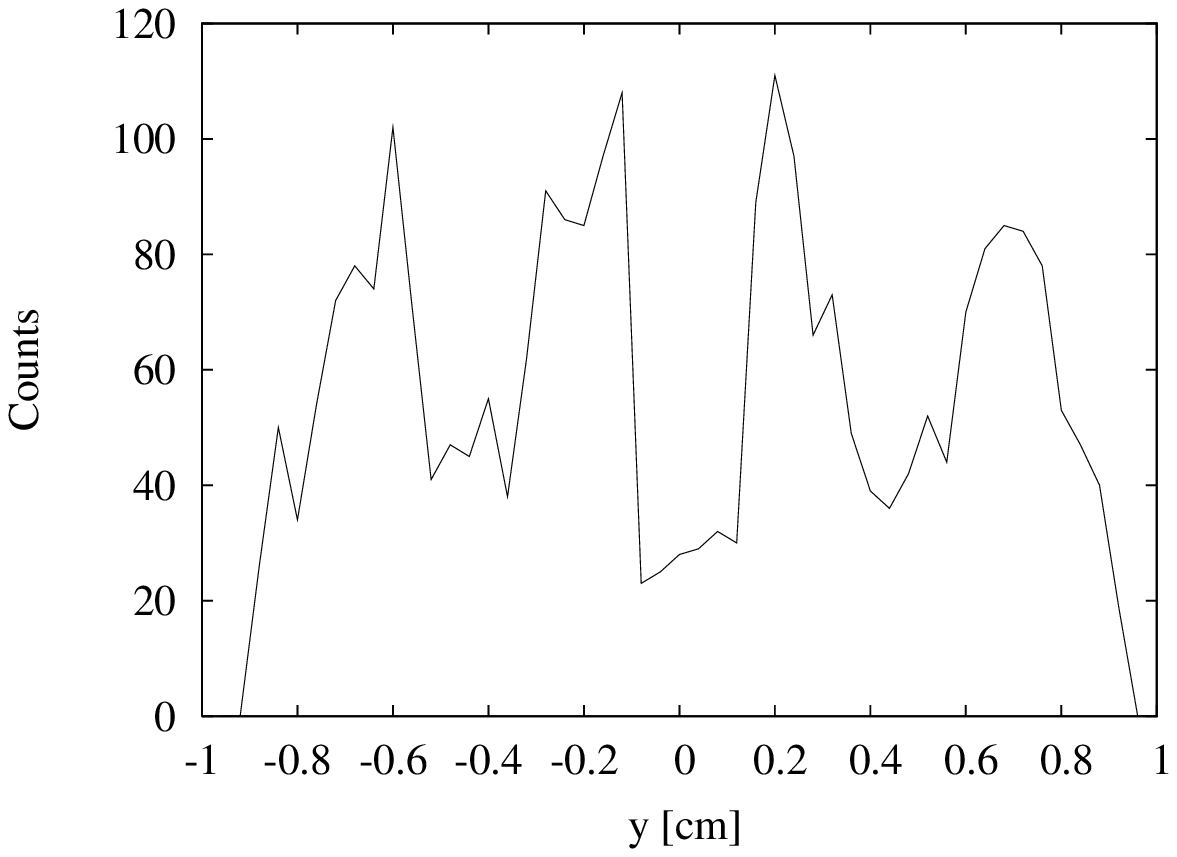}
\caption{}
\label{mono_3_ist}
\end{center}
\end{figure}

\noindent Fig. \ref{mono_3_ist}. Three-dimensional density distribution of the photons in Fig. \ref{mono_3}. Source and screen are in points $s$ and $P2$, respectively. A section taken along the y-axis confirms that even in the center of the spot the radiation distribution is never equal to zero.

\begin{figure}[htbp]
\begin{center}
\includegraphics[width=8.4 cm]{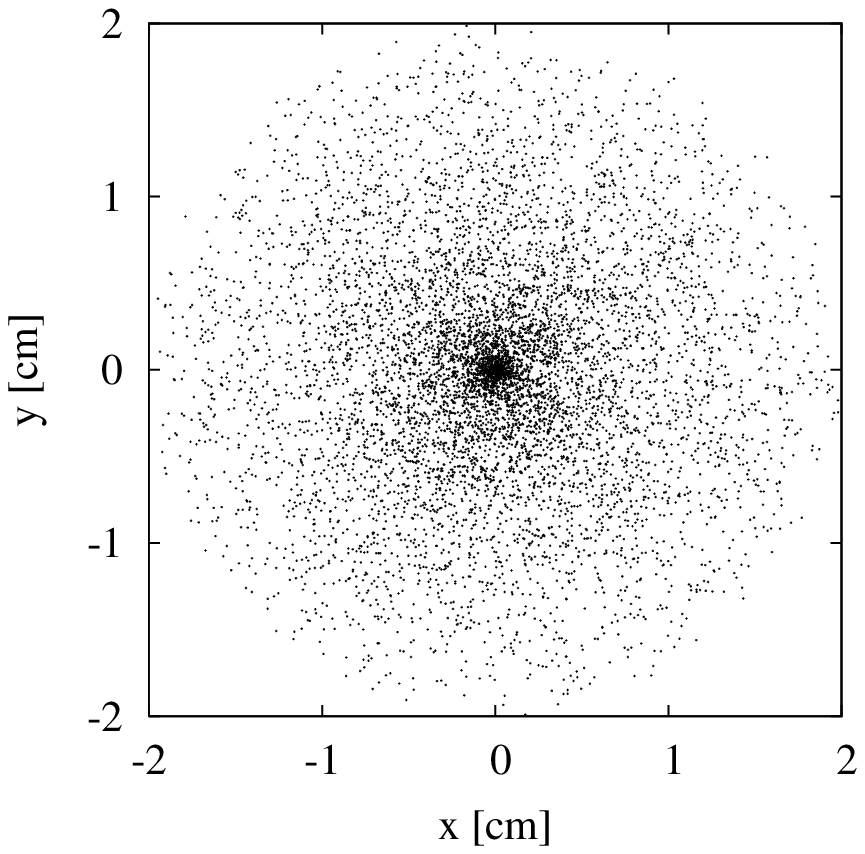}
\caption{}
\label{poli_f}%
\end{center}
\end{figure}

\noindent Fig. \ref{poli_f}. Image of a point source on a screen behind a \emph{cylindrical polycapillary}. The main difference between the polycapillary and monocapillary is that the halo is not homogeneous, Polycapillary parameters: length 10 cm and radius 1 cm. Radius of each capillary is 0.001 cm.

\begin{figure}[htbp]
\begin{center}
\includegraphics[width=8.4 cm]{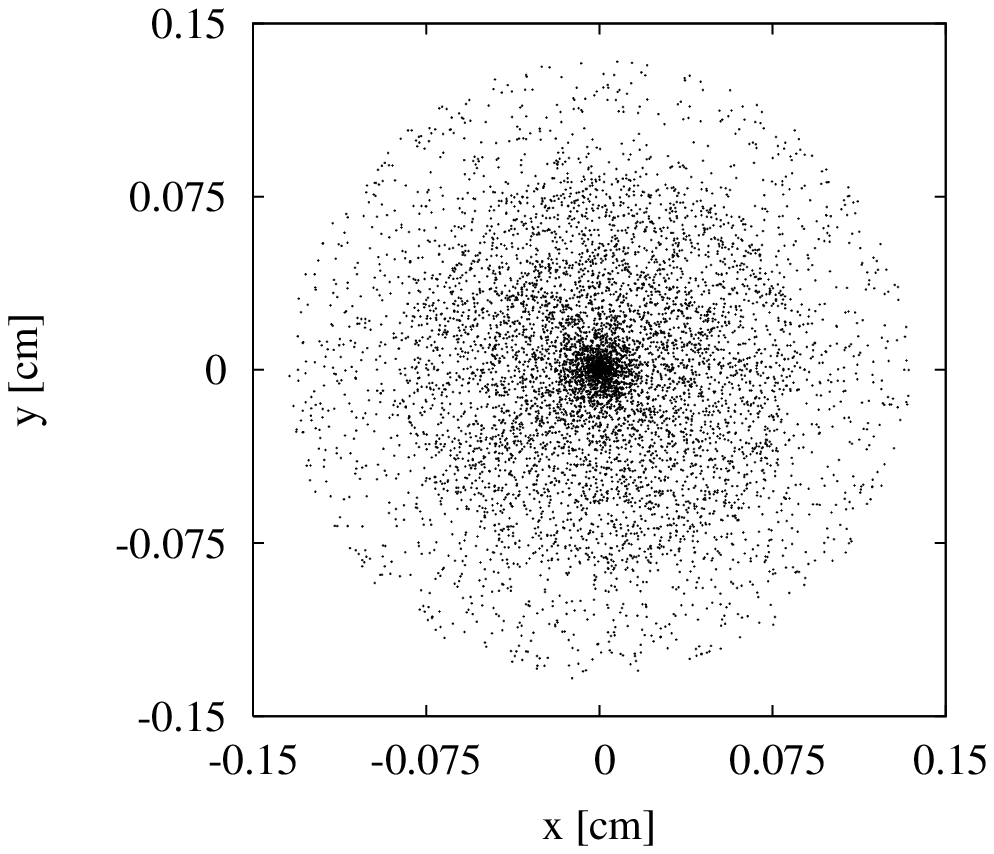}
\caption{}
\label{mono_con_f}%
\end{center}
\end{figure}

\noindent Fig. \ref{mono_con_f}. Image formed by a \emph{conical monocapillary} in the best geometrical conditions with the screen on the focal point (see text for details). Channel parameters: radius at capillary entrance $10^{-1}$ cm; radius at exit $0.8\times10^{-1}$ cm; energy 1 keV; monocapillary length L=10 cm.

\begin{figure}[htbp]
\begin{center}
\includegraphics[width=8.4 cm]{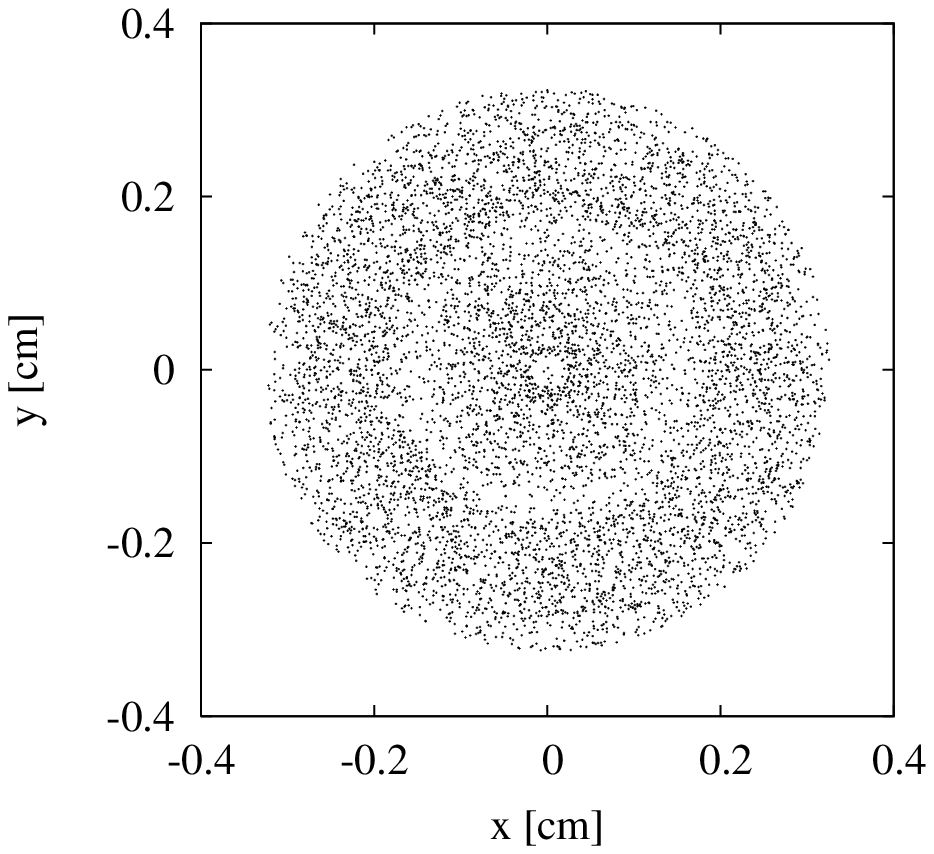}
\caption{}
\label{mono_con_3}%
\end{center}
\end{figure}

\noindent Fig. \ref{mono_con_3}. \emph{Conical monocapillary} output image on a screen located out of focus. Here, the concentric rings have higher intensity than the central spot. Parameters same as in Fig. \ref{mono_con_f}.

\begin{figure}[htbp]
\begin{center}
\includegraphics[width=8.4 cm]{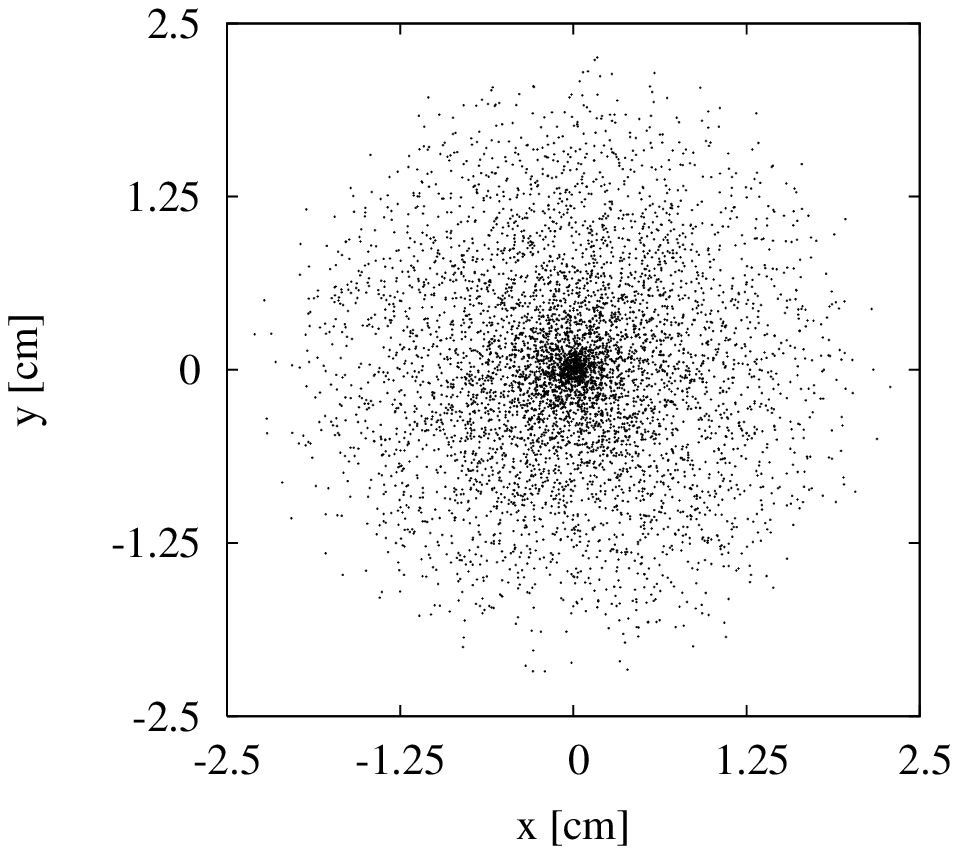}
\caption{}
\label{poli_con}%
\end{center}
\end{figure}

\noindent Fig. \ref{poli_con}. Point-source image on a screen behind a \emph{conical polycapillary}. As in Fig. \ref{poli_f} the main difference is that the halo is not homogeneous. Polycapillary parameters: length = 10 cm, entrance radius 1 cm and exit radius 0.8 cm. Radius of each capillary is $10^{-3}$ cm at entrance and $0.8\times10^{-3}$ cm at exit.

\begin{figure}[htbp]
\begin{center}
\includegraphics[width=8.4 cm]{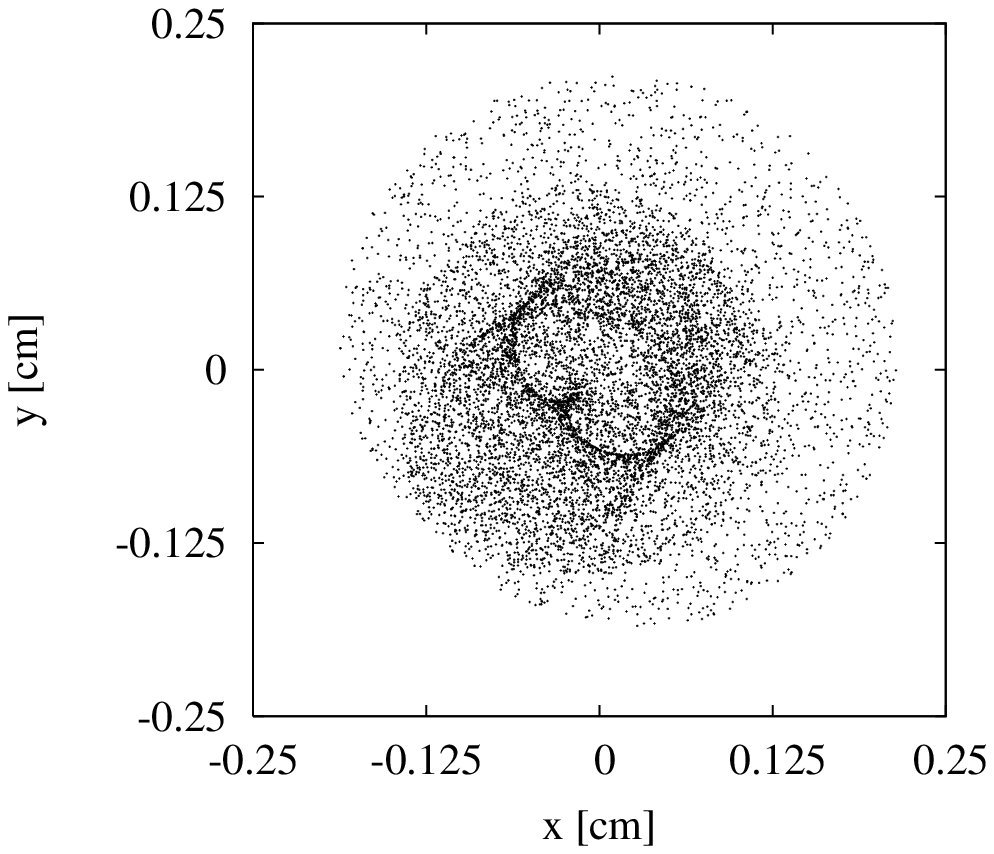}
\caption{}
\label{mono_off}%
\end{center}
\end{figure}

\noindent Fig. \ref{mono_off}. Image due to a \emph{single cylindrical channel}, when point source off-axis. Note that there are no areas without any count. Capillary parameters same as in Fig. \ref{mono_f}. Source is at point $(0.05, 0.05, 3,3)$ cm from entrance plane.

\begin{figure}[htbp]
\begin{center}
\includegraphics[width=8.4 cm]{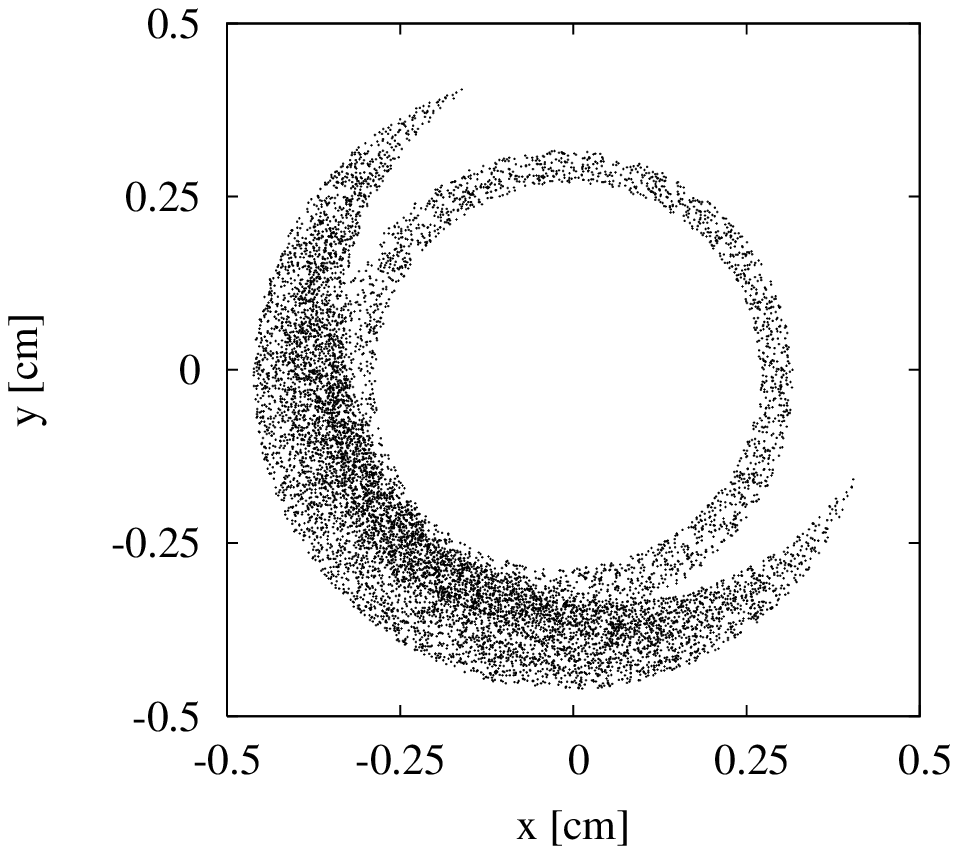}
\caption{}
\label{mono_para}%
\end{center}
\end{figure}

\noindent Fig. \ref{mono_para}. Image with point source at infinite distance (a typical astrophysics source). \emph{Cylindrical monocapillary} parameters same as in Fig. \ref{mono_off}. Source parallel beam has zenithal angle $\alpha=1.3^{\circ}$ and equatorial angle $\delta=45^{\circ}$ (Cf. Fig. \ref{geometry}).

\begin{figure}[htbp]
\begin{center}
\includegraphics[width=8.4 cm]{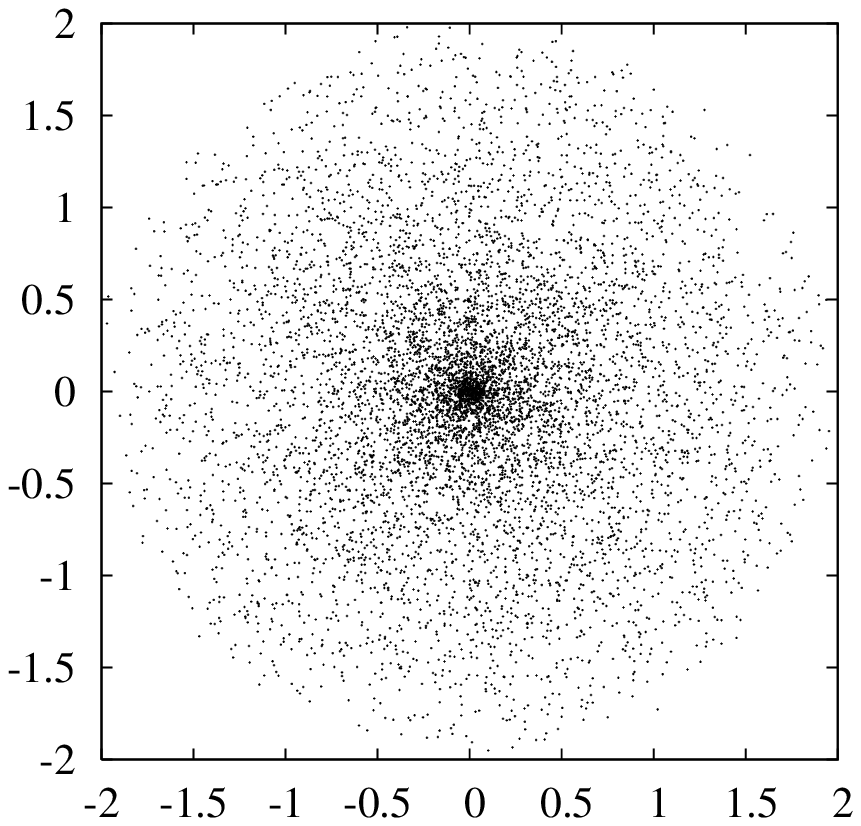}
\caption{}
\label{point}%
\end{center}
\end{figure}

\noindent Fig. \ref{point}. The image formed on the screen by the cylindrical polycapillary, when the `optical point focus' of a Cu tube has been used.

\begin{figure}[htbp]
\begin{center}
\includegraphics[width=8.4 cm]{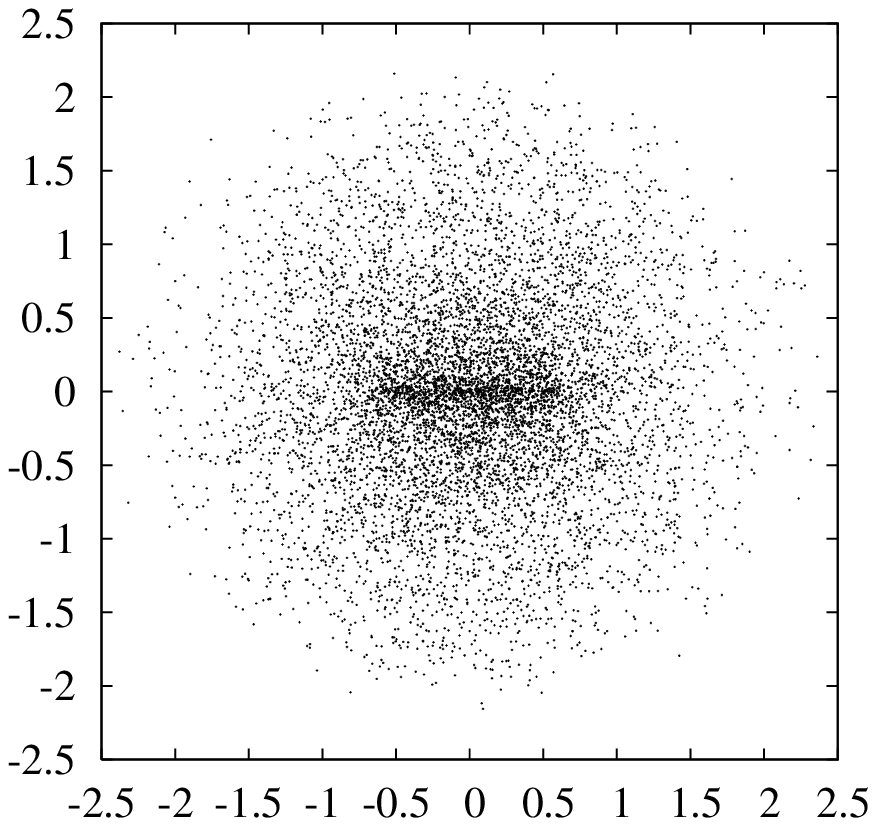}
\caption{}
\label{line}%
\end{center}
\end{figure}

\noindent Fig. \ref{line}. The image behind cylindrical polycapillary sample when the `optical line focus' of a Cu tube has been used.

\newpage

\newpage

\vspace{15cm}

\vspace{15cm}

\vspace{15cm}

\vspace{15cm}

\vspace{15cm}

\vspace{15cm}

\vspace{15cm}

\vspace{15cm}

\vspace{15cm}

\vspace{15cm}

\vspace{15cm}

\vspace{15cm}


\vspace{15cm}


\end{document}